\documentclass[12pt]{revtex4}

\usepackage{graphicx}
\usepackage{amssymb}
\usepackage{times}
\usepackage{amsmath}

\newcommand{\bra}[1]{\langle#1|}
\newcommand{\ket}[1]{|#1\rangle}

\begin{document}

\title{Coherent State Distinguishability in Continuous Variable Quantum Cryptography}

\author{Christian~Weedbrook} \affiliation{Department of Physics,
University of Queensland, St Lucia, Queensland 4072, Australia}

\author{Mile Gu} \affiliation{Department of Physics,
University of Queensland, St Lucia, Queensland 4072, Australia}

\author{Andrew~M.~Lance} \affiliation{Quantum Optics Group, Department
of Physics, Faculty of Science, Australian National University,
ACT 0200, Australia}

\author{Thomas~Symul} \affiliation{Quantum Optics Group, Department
of Physics, Faculty of Science, Australian National University,
ACT 0200, Australia}

\author{Ping~Koy~Lam} \affiliation{Quantum Optics Group, Department
of Physics, Faculty of Science, Australian National University,
ACT 0200, Australia}

\author{Timothy~C.~Ralph} \affiliation{Department of Physics, University of Queensland, St Lucia, Queensland 4072, Australia}

\date{\today}

\begin{abstract}

We use the probability of error as a measure of distinguishability
between two pure and two mixed symmetric coherent states in the
context of continuous variable quantum cryptography. We show that
the two mixed symmetric coherent states (in which the various
components have the same real part) never give an eavesdropper
more information than two pure coherent states.

\end{abstract}

\maketitle

 \setcounter{section}{0}

\section{Introduction}

The security of coherent state continuous variable quantum key
distribution (CV-QKD) \cite{Wee04,Sil02} is fundamentally based on
the inability of an eavesdropper to perfectly distinguish between
non-orthogonal quantum states \cite{Nie00}. In this paper, we look
at how much information a potential eavesdropper can gain when
trying to distinguish between two pure coherent states as opposed
to distinguishing between two mixed coherent states. This is of
particular interest in CV-QKD protocols, such as post-selection
\cite{Sil02}), where it is important to determine if an
eavesdropper obtained more information in the case of
distinguishing between two pure coherent states or distinguishing
between two mixed states.

\section{Probability of Error}

\label{sec-2}

In our analysis, we will use the \textit{probability of error}
($PE$) measure to distinguish between quantum states. We point out
that one could potentially consider other distinguishability
measures such as the Kolmogorov distance, the Bhattacharyya
coefficient and the Shannon distinguishability (for a review of
these measures see \cite{Fuc97}). However, as we shall see the
probability of error measure has a number of useful properties and
can be directly calculated for the quantum states we consider in
our analysis.

We consider the distinguishability between two general quantum
states that are described by the two density matrices
$\hat{\rho}_0$ and $\hat{\rho}_1$. It was originally shown by
Helstrom \cite{Hel76} that the probability of error between these
two density matrices is minimized by performing an optimal
positive operator-valued measure (POVM) \cite{Nie00}. In this
case, the probability of error for the distinguishing between two
general quantum states can be expressed as \cite{Fuc97}
\begin{eqnarray}
PE(\hat{\rho}_0,\hat{\rho}_1) = \frac{1}{2} - \frac{1}{4}
\sum^N_{j=1} |\lambda_j|  = \frac{1}{2} - \frac{1}{4} \rm{Tr}
|\hat{\rho}_0-\hat{\rho}_1| \label{Error_Probability}
\end{eqnarray}
where $PE(\hat{\rho}_0,\hat{\rho}_1) \in [0,1/2]$ and $\lambda_j$
are the eigenvalues of the matrix $\hat{\rho}_0-\hat{\rho}_1$. We
note that when the two states are indistinguishable the
probability of error is $PE=1/2$. On the other hand, in the case
when the two states are completely distinguishable the probability
of error is $PE=0$.

\begin{figure}[!ht]
\begin{center}
\includegraphics[width=12cm]{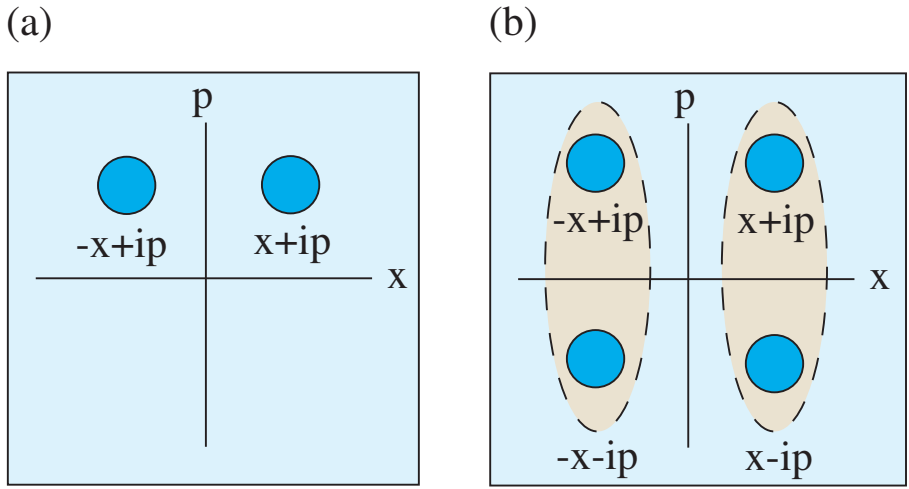}
\caption{Phase space representation of (a) Two pure coherent
states ($\rho_{p_0}$ and $\rho_{p_1}$) and (b) two mixed coherent
states ($\rho_{m_0}$ and $\rho_{m_1}$). Here the dotted lines and
shadings in (b) indicate which of the two coherent states are
mixed.}\label{two_mixed_vs_two_pure_discrete}
\end{center}
\end{figure}

\section{Distinguishing Pure and Mixed Coherent States}
\label{sec-3}

We now look at distinguishing between two coherent states using
the previously defined probability of error. A coherent state is
defined as $\ket{\alpha} = \hat{D}\ket{0}$ where $\hat{D} = {\rm
exp}(\alpha \hat{a}^{\dagger}-\alpha^* \hat{a})$ is the
displacement operator. It is also a minimum uncertainty state and
an eigenstate of the annihilation operator $\hat{a}$, i.e.
\begin{equation}
\hat{a} \ket{\alpha} = \alpha \ket{\alpha}
\end{equation}
where $\alpha$ is the amplitude of the electromagnetic wave
\cite{Ger05}. Any two coherent states $\ket{\alpha}$ and
$\ket{\beta}$ are always non-orthogonal and only approach
orthogonality (i.e. $\bra{\alpha} \beta \rangle \rightarrow 0$)
when $|\alpha-\beta| \gg 1$ where the magnitude is $|\bra{\alpha}
\beta \rangle|^2 = \rm{exp}(-|\alpha-\beta|^2)$. In the following
analysis we will define a coherent state displace in the amplitude
and phase quadratures \cite{Ger05}, by an amount $x$ and $p$
respectively, as $\ket{\alpha}= \ket{x + i p}$. Consequently, we
can write the density operators of two pure coherent states
$\hat{\rho}_{p0}$ and $\hat{\rho}_{p1}$ that we consider here as
\begin{eqnarray}
\hat{\rho}_{p0} &=& |x + ip\rangle \langle x + ip| \\\nonumber
\hat{\rho}_{p1} &=& |-x + ip\rangle \langle -x + ip|
\end{eqnarray}
In our analysis we also consider two mixed coherent states, i.e.
an equally weighted mixture of coherent states mirrored in the
phase quadrature and with both mixtures having the same amplitude
component. The density operators corresponding to these two mixed
states, $\hat{\rho}_{m0}$ and $\hat{\rho}_{m1}$, are defined as

\begin{eqnarray}
\hat{\rho}_{m0} &=& \frac{1}{2} (|x + ip\rangle \langle x + ip| +
|x - ip\rangle \langle x - ip|)
\\\nonumber
\hat{\rho}_{m1} &=& \frac{1}{2}|-x + ip\rangle \langle -x + ip| +
|-x - ip\rangle \langle -x - ip|)
\end{eqnarray}
Figure~(\ref{two_mixed_vs_two_pure_discrete}a) and
Fig.~(\ref{two_mixed_vs_two_pure_discrete}b) give a
two-dimensional phase space illustration of the two pure coherent
states and the two mixed coherent states defined by Eq.~(2) and
Eq.~(3) respectively.

According to Eq.~(\ref{Error_Probability}) we need to determine
the eigenvalues of $\hat{A} = \hat{\rho}_{0} - \hat{\rho}_{1}$ for
both the two pure states and two mixed states. To do this we write
$\hat{A}$ in its matrix representation which can be expanded in
terms of the orthogonal Fock or number states $\ket{n}$ defined as
\cite{Ger05}
\begin{equation}
\ket{n} = \frac{(\hat{a}^{\dag})^n}{\sqrt{n!}} \ket{0}
\end{equation}
where $\hat{a}^{\dag}$ is the creation operator of a harmonic
oscillator and $n \in [0,\infty)$. For example, the coherent state
$\ket{x+ip}$ written in terms of this Fock basis is
\begin{equation}
\ket{x+ip} = e^{-|x+ip|^2/2}\sum \frac{(x+ip)^n}{
\sqrt{n!}}\ket{n}
\end{equation}
Once $\hat{A}$ is written in matrix form we can then numerically
determine its eigenvalues. In this Fock state expansion the inner
product of an arbitrary coherent state with a Fock state is given
by
\begin{eqnarray}
\langle n|\pm x \pm ip \rangle = \frac{(\pm x \pm
ip)^n}{\sqrt{n!}}{\rm exp} (-\frac{1}{2}(x^2 + p^2))\\
\langle \pm x \pm ip|m \rangle = \frac{(\pm x \mp
ip)^m}{\sqrt{m!}}{\rm exp} (-\frac{1}{2}(x^2 + p^2))
\end{eqnarray}
where $\ket{n}$ and $\ket{m}$ are Fock states. Calculating the
general matrix coefficients for the case of the two pure coherent
states we
obtain 
\begin{eqnarray}\label{matrices_pure}
\langle n|A|m\rangle_{pure} &=& \frac{{\rm exp}(-x^2 -
p^2)}{\sqrt{n!m!}}[(x+ip)^n(x-ip)^m - (-x+ip)^n(-x-ip)^m]
\end{eqnarray}
Similarly for the two mixed state case we find
\begin{eqnarray}\label{matrices_mixed}\nonumber
\langle n|A|m\rangle_{mixed} &=& \frac{{\rm exp}(-x^2 -
p^2)}{2\sqrt{n!m!}}[(x+ip)^n(x-ip)^m + (x-ip)^n(x+ip)^m\\
&-& (-x+ip)^n(-x-ip)^m - (-x-ip)^n(-x+ip)^m]
\end{eqnarray}
Numerically we calculate the eigenvalues of
Eq.~(\ref{matrices_pure}) and Eq.~(\ref{matrices_mixed}) up to
certain values of $n$ and $m$. Then according to
Eq.~(\ref{Error_Probability}) this will give us the probability of
error in distinguishing between two quantum states. Now having
numerically calculated $PE$ we would like to interpret this in
terms of the information gained from using the distinguishing
measure.

\section{Shannon Information}

\begin{figure}[!ht]
\begin{center}
\includegraphics[width=8cm]{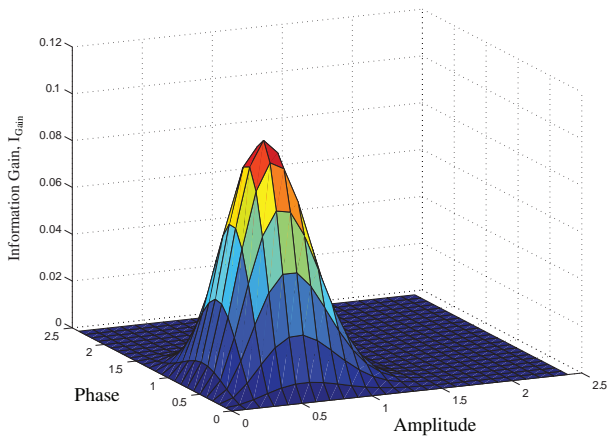}
\caption{The difference in information rates between the two pure
states and two mixed states in terms of the amplitude and phase
quadratures. Here $I_{gain} = I(\rho_{p_0},\rho_{p_1})-
I(\rho_{m_0},\rho_{m_1})$. In this case the two mixed states never
give more information than two pure states.}\label{3D_plot}
\end{center}
\end{figure}

In the context of CV-QKD it is important to determine how much
Shannon information an eavesdropper can obtain by distinguishing
between two (pure or mixed) quantum states. The information
obtained by distinguishing between two states can be calculated
using the Shannon information formula for a binary symmetric
channel \cite{Sha48}
\begin{eqnarray}\label{shannon_formula}
I = 1 + PE{\rm log_{2}}PE + (1 - PE){\rm log_{2}}(1 - PE).
\end{eqnarray}
Figure~(\ref{3D_plot}) shows the difference between the Shannon
information obtained by distinguishing between two coherent states
$I(\rho_{p_0},\rho_{p_1})$ compared with distinguishing between
two mixed states $I(\rho_{m_0},\rho_{m_1})$. This information
difference is defined as $I_{gain} = I(\rho_{p_0},\rho_{p_1})-
I(\rho_{m_0},\rho_{m_1})$. Figure~(\ref{3D_plot}) plots $I_{gain}$
in terms of the amplitude and phase quadrature displacements of
the pure and mixed states as defined in Eq.~(3) and Eq.~(4),
respectively. Here we have expanded up to $50$ Fock states, i.e.
$n = m = 50$ in our numerical analysis.

There are two main features of this plot. Firstly, we notice that,
given our distinguishability measure and initial configuration of
coherent states in phase space, two mixed states never give more
information than two pure state, i.e.
%
$I(\hat{\rho}_{m0},\hat{\rho}_{m1}) \leq
I(\hat{\rho}_{p0},\hat{\rho}_{p1})$.
%
Secondly, there is a flat region where the information gain is
zero, i.e. the information from distinguishing between two mixed
states is the same as that of two pure states. This means as we
move the states further and further apart in the amplitude
quadrature (while keeping the phase quadrature fixed), the
probability of error tends to zero and hence an information gain
of zero. The same result occurs when the amplitude quadrature is
fixed while varying the phase quadrature. This is somewhat
surprising because the more separated two mixed states become the
more indistinguishable they are and consequently less information
can be extracted. But in this case what it is telling us is that
at some point the two mixed states start ``behaving" like two pure
states.

\section{Conclusion}
\label{sec-5}

In conclusion, we have shown that a continuous variable quantum
key distribution protocol where an eavesdropper needs to
distinguish between two pure coherent states, rather than two
mixed coherent states (where the various mixtures have the same
amplitude component), the eavesdropper will never get more
information from the two mixed coherent states. We showed this
using the probability of error as the distinguishability measure
along with the Shannon information formula.

We would like to thank the Australian Research Council for
support.

\end{document}